\documentclass[twocolumn,pra,superscriptaddress,english]{revtex4}
\usepackage{amssymb}
\usepackage{amsmath}
\usepackage{graphicx}
\usepackage{subfigure}
\usepackage{natbib}
\usepackage{epsfig}
\usepackage{amsfonts}
\usepackage{mathrsfs}
\usepackage{color}
\usepackage{bbm}
\usepackage{bm}
\usepackage{hyperref}
\usepackage[toc,page,title,titletoc,header]{appendix}

\begin{document}


\global\long\def\id{\mathbbm{1}}
\global\long\def\ui{\mathbbm{i}}
\global\long\def\ud{\mathrm{d}}

\title{Characterizing topological phases by quantum quenches: A general theory}

\author{Long Zhang}
\thanks{These authors contribute equally to this work.}
\affiliation{International Center for Quantum Materials and School of Physics, Peking University, Beijing 100871, China}
\affiliation{Collaborative Innovation Center of Quantum Matter, Beijing 100871, China.}

\author{Lin Zhang}
\thanks{These authors contribute equally to this work.}
\affiliation{International Center for Quantum Materials and School of Physics, Peking University, Beijing 100871, China}
\affiliation{Collaborative Innovation Center of Quantum Matter, Beijing 100871, China.}

\author{Xiong-Jun Liu
\footnote{Correspondence author: xiongjunliu@pku.edu.cn}}
\affiliation{International Center for Quantum Materials and School of Physics, Peking University, Beijing 100871, China}
\affiliation{Collaborative Innovation Center of Quantum Matter, Beijing 100871, China.}
\affiliation{Beijing Academy of Quantum Information Science, Beijing 100193, China}
\affiliation{CAS Center for Excellence in Topological Quantum Computation, University of Chinese Academy of Sciences, Beijing 100190, China}
\affiliation{Institute for Quantum Science and Engineering and Department of Physics, Southern University of Science and Technology, Shenzhen 518055, China}


\begin{abstract}
We investigate a generic dynamical theory to characterize topological quantum phases by quantum quenches, 
and study the emergent topology of quantum dynamics when the quenches start from a deep or shallow trivial phase to topological regimes. 
Two dynamical schemes are examined: One is to characterize topological phases via quantum dynamics induced by 
a single quench along an arbitrary axis, and the other applies a sequence of quenches with respect to all (pseudo)spin axes.
These two schemes are both built on the so-called dynamical bulk-surface correspondence, which shows that the $d$-dimensional ($d$D)
topological phases with integer invariants can be characterized by the dynamical topological pattern emerging on $(d-1)$D band inversion surfaces (BISs).
We show that the first dynamical scheme works for both deep and shallow quenches, the latter of which is initialized in an incompletely polarized trivial phase. 
For the second scheme, however, when the initial phase for the quench study varies from the deep trivial (fully polarized) regime to shallow trivial (incompletely polarized) regime, 
a new dynamical topological transition, associated with topological charges crossing BISs, is predicted in quench dynamics. 
A generic criterion of the dynamical topological transition is precisely obtained. Above the criterion (deep quench regime), 
quantum dynamics on BISs can well characterize the topology of the post-quench Hamiltonian.
Below the criterion (shallow quench regime), the quench dynamics may depict a new dynamical topology; 
the post-quench topology can be characterized by
the emergent topological invariant plus the total charges moving outside the region enclosed by BISs. 
We illustrate our results by numerically calculating the 2D quantum anomalous Hall model, 
which has been realized in ultracold atoms. This work broadens the way to classify topological phases by non-equilibrium quantum dynamics, 
and has feasibility for experimental realization.

\end{abstract}

\maketitle

\section{Introduction}
Since the discovery of the quantum Hall effect~\cite{Klitzing1980,Tsui1982}, topological quantum phases have been a hot research area in condensed matter physics.
The topological quantum phase is beyond the Landau-Ginzburg-Wilson symmetry-breaking paradigm~\cite{Landau1999}, and 
can be characterized by the global bulk topological invariant defined in the ground state at equilibrium~\cite{Hasan2010,Qi2011}.
Among the exotic features, a most fundamental phenomenon of topological quantum phases is the bulk-boundary correspondence which relates the protected boundary modes to bulk topology.
The bulk-boundary correspondence allows for the detection of topological insulators~\cite{Konig2007,Hsieh2008,Xia2009,Chang2013}
and semimetals~\cite{Xu2015, Lv2015} through transport measurement or angle-resolved photoemission spectroscopy, which can directly resolve boundary modes in solid state systems. 

On the other hand, being a clean environment and of full controllability~\cite{Bloch2008}, ultracold atoms provide an ideal platform to realize novel topological models and explore exotic topological physics~\cite{Atala2013,Liu2013,Aidelsburger2013,Miyake2013,Jotzu2014,Aidelsburger2015,Wu2016,Flaschner2016,Baozong2018,Sun2017,Song2018}. In particular, compared with solid state systems, 
realization in ultracold atoms also facilitates the study of non-equilibrium quantum dynamics in topological phases by quantum quenches~\cite{Song2018,DQPT1,DQPT2,WYi2018,Flaschner2018,Song2018b}. 
For Chern insulators, the unitary evolution after quench does not change the topology of the many-body state. Accordingly, if the system is initialized in the trivial phase, 
the many-body state keeps in the trivial phase during the unitary evolution even when the post-quench target phase is topologically nontrivial~\cite{unitary1,unitary2,unitary3,unitary4}. 
This implies that the bulk-boundary correspondence for the equilibrium topological phases may not be satisfied in the dynamical regime. 
Nevertheless, the quench-induced evolution may carry the information of the topology of the post-quench phase. 
In particular, a novel {\it dynamical bulk-surface correspondence} was proven recently and applies to the generic $d$-dimensional ($d$D) topological phase~\cite{Zhanglin2018}, 
showing that the equilibrium bulk topology for a generic $d$D topological phase universally corresponds to the dynamical topology emerging in ($d-1$)D momentum subspaces, 
dubbed {\it band inversion surfaces} (BISs). Similar to the bulk-boundary correspondence for equilibrium topological phases in the real space, 
the dynamical bulk-surface correspondence is of broad applicability and opens up generic non-equilibrium characterization or classification for topological phases~\cite{Zhanglong2018,Yang2018,Gong2018,McGinley2018a,McGinley2018b,Yu2019}, with novel experimental progresses having been made recently~\cite{Sun2018,Tarnowski2019,Wang2019,Yi2019,TopoChargeExp}.

Building on the dynamical bulk-surface correspondence, the non-equilibrium characterization theory~\cite{Zhanglin2018} provides new schemes 
with explicit advantages over equilibrium schemes to measure the bulk topology of equilibrium topological quantum phases. 
The main advantages are listed as follows:
(i) The dynamical bulk-surface correspondence simplifies the topological characterization to lower-dimensional invariants on BISs.
(ii) Quench dynamics is resonant and nontrivial only on BISs, so the characterization is easily resolved in experiment.
(iii) Short-term quench dynamics provides enough information to characterize the post-quench topology, which
is hardly affected by detrimental effects like thermal effects, thus leading to high-precision measurement of topological phase diagrams, as confirmed in experiment~\cite{Sun2018}.
Unlike the scheme in Ref.~\cite{Zhanglin2018}, which employs the quench along a single (pseudo)spin axis but measures spin polarizations 
in all directions to extract the topology, an alternative dynamical classification scheme was further introduced to simplify the measurement~\cite{Zhanglong2018}.
In this scheme, the topology was proposed to be characterized by topological charges, which can be precisely detected 
by measuring only a single spin component after a sequence of quenches along all spin quantization axes~\cite{Zhanglong2018}.
These two schemes~\cite{Zhanglin2018,Zhanglong2018} 
have been demonstrated, respectively, in a diamond nitrogen-vacancy center spin system~\cite{Wang2019} and
with ultracold $^{87}$Rb bosons trapped in an optical Raman lattice~\cite{TopoChargeExp}.
Moreover, the dynamical bulk-surface correspondence has been further applied to an interacting system~\cite{Zhanglong2019}, in which
the correlated quench dynamics also exhibits robust topological structure on BISs despite of interaction induced dephasing and heating.
Non-equilibrium characterization has also been considered in non-Hermitian systems~\cite{Zhou2018,Jiang2018,Xue2019,Ghatak2019}.

Previous schemes~\cite{Zhanglin2018,Zhanglong2018} have mainly considered deep quenches, 
which initialize a completely polarized trivial state and induces non-equilibrium dynamics by quenching the system to a topologically nontrivial phase.
In this work, we extend the non-equilibrium characterization theory to the more generic
case that the quench starts from a generic trivial phase with incomplete polarization, and predict interesting new physics.
We call such a quench the ``shallow'' quench.  We shall examine the validity of dynamical bulk-surface correspondence
in the dynamical characterization via a single or a sequence of shallow quenches.
On one hand, using a local transformation, we find that the dynamical classification via a single quench can be straightforwardly applied to the incompletely polarized case
except that the so-called {\it dynamical} band inversion surfaces (dBISs) should take the place of BISs.
Although dBISs depend on the initial state and are generally not equal to BISs,
we demonstrate that the post-quench quantum phase can be characterized by the topological pattern emerging on dBISs.
On the other hand, we also consider the generalization of the dynamical classification via quenches along all spin polarization axes.
Interestingly, we find that the validity of this scheme depends on the overall polarization of the initial trivial phase and the dimensionality of the system.
This result indicates that there exists a transition of the emergent topology exhibited by quench dynamics.
We shall show that the dynamical topological transition happens when a {\it dynamical} topological charge moves across the BIS,
giving rise to a change of the total charges enclosed by BISs.

The remaining is organized as follows. In Sec.~\ref{Sec2}, we review the dynamical classification scheme that employs a single deep quench.
In Sec.~\ref{Sec3}, we generalized the non-equilibrium theory to the shallow quench case. We
demonstrate the generalized theory by a local transformation and further illustrate it by numerically calculating the 2D quantum anomalous Hall (QAH) model.
In Sec.~\ref{Sec4}, we consider the dynamical classification scheme via a sequence of quenches.
We discuss the validity condition and the emergent dynamical topological transition. We also use the 2D QAH model as an illustration.
Brief discussion and summary are presented in Sec.~\ref{Sec5} and more details can be found in the Appendixes.

\section{Classification by a deep quench}\label{Sec2}
This section briefly reviews the dynamical classification theory we developed in
Refs.~\cite{Zhanglin2018,Zhanglong2018}, in which the
quench process is set from a deep trivial to a topological regime.
This non-equilibrium theory can be applied to generic $d$-dimensional ($d$D) gapped topological phases (including insulators
and superconductors) that are classified by integer invariants in the Altland-Zirnbauer (AZ) symmetry
classes~\cite{AZ1997,Schnyder2008,Kitaev2009,Chiu2016}.

The basic Hamiltonian can be written in the elementary representation matrices of the Clifford algebra~\cite{Morimoto2013, Chiu2013} as
\begin{equation}\label{model}
\mathcal{H}(\mathbf{k})={\bf h}(\mathbf{k})\cdot{\bm\gamma}=\sum_{i=0}^{d}h_{i}(\mathbf{k})\gamma_{i},
\end{equation}
where the $\gamma$ matrices obey anti-commutation relations $\left\{ \gamma_{i},\gamma_{j}\right\}=2\delta_{ij}\id$,
with $i,j=0,1,\dots,d$, and ${\bf h}(\mathbf{k})$ describes a $(d+1)$D Zeeman field depending
on the Bloch momentum ${\bold k}$ in the Brillouin zone (BZ).
We emphasize that the $\gamma$ matrices are constructed to satisfy the trace property ${\rm Tr}[\prod_{j=0}^d\gamma_j]=(-2\ui)^n$
for even $d=2n$ or ${\rm Tr}[\gamma\prod_{j=0}^d\gamma_j]=(-2\ui)^n$ for odd $d=2n-1$, with $\gamma=\ui^n\prod_{j=0}^d\gamma_j$
being the chiral matrix. For example, in two dimensions we should have ${\rm Tr}[\gamma_0\gamma_1\gamma_2]=-2\ui$; if $\gamma_0=\sigma_z$,
one should set $\gamma_1=\sigma_y$ and $\gamma_2=\sigma_x$.
Here the dimensionality of $\gamma$ matrices reads $n_d=2^n$,
which is the minimal requirement to open a topological gap for the $d$D topological phase~\cite{Chiu2013}.
Although it is formulated with the basic Hamiltonian, the classification theory is applicable to any generic multiband
model of the $d$D phase that can be decomposed into a direct sum of the above basic Hamiltonian~\cite{Zhanglin2018}.
The topology of the Hamiltonian $\mathcal{H}(\mathbf{k})$ is
classified by $d$D winding
number (if $d=2n-1$) or the $n$-th Chern
number ($d=2n$). The winding (or Chern) number is determined by the unit vector field
$\hat{\bf h}({\bf k})\equiv{\bf h}({\bf k})/|{\bf h}({\bf k})|$, which
is a mapping from the BZ torus $T^d$ to the $d$D spherical surface $S^d$.
Geometrically speaking,
the topological number counts the times that the
mapping covers $S^d$~\cite{Fruchart2013}.

Without loss of generality, we choose $h_{0}(\mathbf{k})$ in the Hamiltonian (\ref{model}) to characterize the band structure,
with the remaining components $h_{i}(\mathbf{k})$ ($i=1,2,...,d$) depicting the inter-band coupling. The $d$-component field
is denoted as spin-orbit (SO) field  ${\bf h}_{\mathrm{so}}(\mathbf{k})\equiv(h_{1},\dots,h_{d})$, corresponding to
the superconducting pairing in topological superconductors.
Without the SO field, band crossing would occur on $(d-1)$D surfaces, called the band inversion surfaces (BISs), defined by
\begin{equation}\label{BIS_definition}
\mathrm{BIS}\equiv\{\mathbf{k}\vert h_{0}(\mathbf{k})=0\}.
\end{equation}
In Ref.~\cite{Zhanglin2018}, we have demonstrated that
the topological characterization by a $d$D topological number can reduce to a
$(d-1)$D invariant defined on BISs, i.e.
the winding of ${\bf h}_{\mathrm{so}}(\mathbf{k})$, which is given by
\begin{equation}\label{wd_s}
{\cal W}=\sum_{j}\frac{\Gamma(d/2)}{2\pi^{d/2}}
\frac{1}{\left(d-1\right)!}\int_{\mathrm{BIS}_j}\hat{\bf h}_{\rm so}(\ud\hat{\bf h}_{\rm so})^{d-1},
\end{equation}
where $\Gamma(x)$ denotes the Gamma function,
$\hat{\bf h}_{\rm so}\equiv{\bf h}_{\rm so}/|{\bf h}_{\rm so}|$ is the unit SO field,
and the summation is taken over all the BISs.

\begin{figure}
\includegraphics[width=0.3\textwidth]{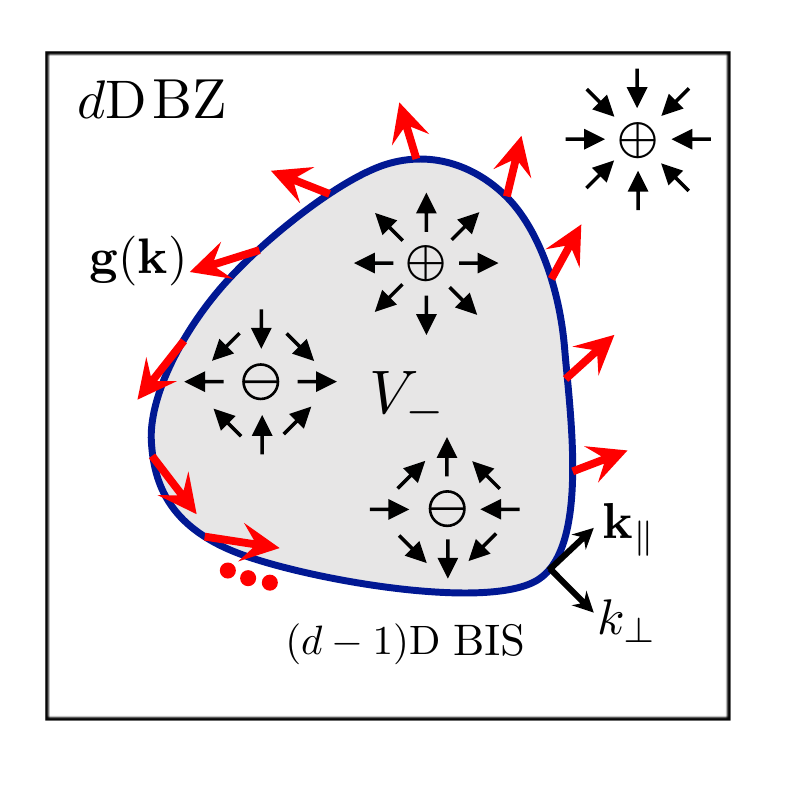}
\caption{Dynamical bulk-surface correspondence and topological charges.
The $d$-dimensional topological phases classified by integer invariants can be characterized by the
winding of the dynamical directional derivative field ${\bf g}({\bf k})$ (red arrows) on the $(d-1)$-dimensional BIS, which
counts the total topological monopole charges in the region $V_-$ (gray) enclosed by BISs with $h_0({\bf k})<0$.
Here the momentum ${\bf k}$ is decomposed into $(k_\perp, {\bf k}_\parallel)$ with $k_\perp$ being perpendicular
to the BIS and pointing to the region with $h_0({\bf k})>0$.
}\label{fig1}
\end{figure}

\subsection{Dynamical bulk-surface correspondence}\label{dbsc}

We present the non-equilibrium characterization by the topological pattern emerging on BISs.
We consider the quench along $\gamma_0$ axis, and write  $h_0({\bf k})=m_0+\nu_0({\bf k})$, where $\nu_0({\bf k})$
denotes the momentum-dependent part and $m_0$ is a constant magnetization. A deep quench then corresponds to setting
$m_0\to\infty$ for $t<0$ and a value in the topological regime for $t\geq0$, which initializes a fully polarized state along the $\gamma_0$ axis.
The time-averaged spin textures are given by
\begin{equation}\label{averages}
\overline{\left\langle \gamma_{i}\right\rangle}=\lim_{T\to\infty}\frac{1}{T}\int_{0}^{T}\ud t\,\mathrm{Tr}\left[\rho_{0}
e^{\ui\mathcal{H}_{\rm post}t}\gamma_{i}e^{-\ui\mathcal{H}_{\rm post}t}\right],
\end{equation}
where $\rho_0$ is the density of matrix of the initial state and $\mathcal{H}_{\rm post}$ denotes the post-quench Hamiltonian.
Due to $e^{\ui {\cal H}t}=\cos(Et)+\ui\sin(Et){\cal H}/E$, with $E^2=\sum_{i=0}^d h_i^2$,
we have
\begin{align}\label{gammai0}
\overline{\langle\gamma_{i}({\bf k})\rangle}=-h_i({\bf k})h_0({\bf k})/E^2({\bf k})
\end{align}
for a deep quench.

On the BIS, the vector field
${\bf h}(\mathbf{k})={\bf h}_{\rm so}(\mathbf{k})$ is perpendicular to initial spin polarization, which
induces spin procession in the plane perpendicular to ${\bf h}_{\rm so}$. As a result, the
time averages $\overline{\left\langle \gamma_{i}\right\rangle}$ should vanish right on BISs for all components.
The dynamical characterization of BISs is thus
\begin{align}\label{BIS0}
\overline{\left\langle \gamma_{i}({\bf k})\right\rangle}=0,\quad{\rm for}\,\,i=0,1,\cdots,d.
\end{align}
From Eq.~(\ref{gammai0}), one can see that this dynamical characterization is equivalent to the definition (\ref{BIS_definition}) in a deep quench.
We further define a dynamical directional derivative field ${\bf g}({\bf k})=(g_1,g_2,\cdots,g_d)$,
whose components are given by
\begin{align}\label{gi}
g_i({\bf k})=-\frac{1}{{\cal N}_k}\partial_{k_\perp}\overline{\left\langle \gamma_{i}\right\rangle},
\end{align}
where $k_\perp$ denotes the momentum perpendicular to BIS,
and ${\cal N}_k$ is the normalization factor.
It can be shown that on the BIS ${\bf g}({\bf k})=\hat{\bf h}_{\rm so}({\bf k})$. Thus the topological invariant
defined in Eq.~(\ref{wd_s}) can take a dynamical form
\begin{equation}\label{wd}
{\cal W}_{\rm d}=\sum_{j}\frac{\Gamma(d/2)}{2\pi^{d/2}}
\frac{1}{\left(d-1\right)!}\int_{\mathrm{BIS}_j}{\bf g}({\bf k})[\ud{\bf g}({\bf k})]^{d-1}.
\end{equation}
The result manifests a highly nontrivial dynamical bulk-surface correspondence (Fig.~\ref{fig1}),
which can be directly measured from the dynamical
spin-polarization patterns emerging on BISs.
Since the quantum spin dynamics is resonant only on BISs, the dynamical
bulk-surface correspondence can be well resolved in experiments.

\subsection{Topological charges}
The band topology can also be viewed from the picture of topological charges~\cite{Zhanglong2018}.
Analogous to the magnetic field, one can consider the monopoles of SO field, and
the topological invariant is thus the flux through BISs.
It is easily seen that the topological charges are located at nodes of ${\bf h}_{\mathrm{so}}({\bf k})$, i.e. the momenta where ${\bf h}_{\mathrm{so}}({\bf k})=0$, such that
the gap is closed and then reopened as a charge passes through a BIS.
The winding of ${\bf h}_{\rm so}$ then
counts the total charges enclosed by BISs.
With this picture we obtain that the topological invariant is given by the summation of monopole charges enclosed
by BISs in the region $V_{-}$ with $h_0({\bf k})<0$~\cite{Zhanglong2018}
\begin{align}\label{topo_invar}
{\cal W}&=\sum_{n\in V_{-}}{\cal C}_n.
\end{align}
Here the topological charge ${\cal C}_n$ characterizes the winding of SO field around the $n$-th monopole. In the typical
case that ${\bf h}_{\mathrm{so}}({\bf k})$ is linear when approaching the monopole, the charge is simplified as
\begin{align}
{\cal C}_n={\rm sgn}[J_{{\bf h}_{\rm so}}({\bm \varrho}_n)],
\end{align}
where $J_{{\bf h}_{\rm so}}({\bf k})\equiv \det\left[(\partial h_{\rm so,i}/\partial k_j)\right]$ is Jacobian determinant.

According to Eq.~(\ref{gammai0}), the location of a monopole charge can be dynamically found by
$\overline{\left\langle \gamma_{i}({\bf k})\right\rangle }=0$ for all $i\neq0$ but
$\overline{\left\langle \gamma_{0}({\bf k})\right\rangle }\neq0$. Furthermore,
near one monopole ${\bf k}={\bm \varrho}_n$,
the time-averaged spin texture
$\overline{\left\langle \gamma_{i}({\bf k})\right\rangle}\big\vert_{{\bf k}\to{\bm \varrho}_n}=-h_{i}(\bold k)/h_{0}({\bm \varrho}_n).$
Hence, we define a dynamical spin-texture field ${\bm \Theta}({\bf k})$ to characterize the charge,
with its components being
\begin{equation}\label{thetai}
\Theta_i({\bf k})\equiv-\frac{{\rm sgn}[h_0({\bf k})]}{{\cal N}_{\bf k}}\overline{\left\langle \gamma_{i}({\bf k})\right\rangle},
\end{equation}
where ${\cal N}_{\bf k}$ is the normalization factor. Thus near one monopole, the dynamical field
$\Theta_i({\bf k})\big\vert_{{\bf k}\to{\bm \varrho}_n}={h}_{{\rm so},i}({\bf k})$. The winding of SO field then reduces to that
of the $\Theta_i({\bf k})$ field, and in the linear case we reach
\begin{equation}
{\cal C}_n={\rm sgn}[J_{{\bm \Theta}}({\bm \varrho}_n)].
\end{equation}

\section{Generic scheme for a shallow quench}~\label{Sec3}
In this section, we generalize our dynamical classification theory to a more generic situation
that the quench starts from a shallow trivial regime.
Such a generalization has practical benefits in experimental realization, where only finite magnetization
can be generated.

\subsection{The projection approach}~\label{Sec3A}
The shallow quench initializes an incompletely polarized state, i.e., at $t=0$,
$\langle\gamma_0({\bf k})\rangle=1$ (or $-1$) does not hold for all ${\bf k}$ but
$\langle\gamma_0({\bf k})\rangle>0$ (or $<0$) does.
We still make use of the time-averaged spin textures defined in Eq.~(\ref{averages}), which in general gives
\begin{align}\label{gammai}
\overline{\langle\gamma_{i}({\bf k})\rangle}=h_i({\bf k}){\rm Tr}[\rho_0{\cal H}_{\rm post}]/E^2({\bf k}).
\end{align}
We still use the momenta with vanishing spin polarizations to define a $(d-1)$D hypersurface, which
is generally not equal to the definition in Eq.~(\ref{BIS_definition}). Hence we name it the {\it dynamical} band inversion surfaces (dBISs):
\begin{align}\label{dBIS}
 \mathrm{dBIS}\equiv\{\mathbf{k}\vert{\rm Tr}[\rho_0(\mathbf{k}){\cal H}_{\rm post}(\mathbf{k})]=0\}.
\end{align}
We define the direction $k_{\perp}$ as being perpendicular to the
contour ${\rm Tr}[\rho_0(\mathbf{k}){\cal H}_{\rm post}(\mathbf{k})]$. The directional derivative on the dBIS then reads
\begin{align}\label{dgamma}
\partial_{k_{\perp}}\overline{\langle\gamma_i\rangle}=\lim_{k_{\perp}\to0}\frac{1}{2k_{\perp}}\frac{h_i+\mathcal{O}(k_{\perp})}
{E^{2}+\mathcal{O}(k_{\perp})}\cdot 2k_{\perp}=\frac{h_i}{E^{2}}.
\end{align}
Hence the field $\partial_{k_\perp}\overline{\langle\boldsymbol{\gamma}\rangle}$ is generally not a $d$-vector, and its winding on dBISs is not well-defined.
Here we propose a projection approach, for which a projected directional derivative field
${\bf g}_{\rm proj}({\bf k})=(g_1,g_2,\cdots,g_d)$ is defined, with
$g_i=-\partial_{k_\perp}\overline{\left\langle \gamma_{i}\right\rangle}/{\cal N}_k$.
From Eq.~(\ref{dgamma}), one can see that the projected field ${\bf g}_{\rm proj}({\bf k})$ characterizes the SO field ${\bf h}_{\rm so}({\bf k})$, and its winding counts the monopole charges enclosed by dBISs.
Note that in the fully polarized case, $\partial_{k_{\perp}}\overline{\left\langle \gamma_{0}\right\rangle}=0$ on dBISs and the projection approach reduces to
the dynamical bulk-surface correspondence in Sec.~\ref{dbsc}. In the following,
we shall show that the topological pattern of the projected directional derivative field ${\bf g}_{\rm proj}({\bf k})$ indeed characterizes the topology of post-quench Hamiltonian.

\subsection{Local rotation and auxiliary Hamiltonian}~\label{Sec3B}

To prove the validity of the projection approach, we make use of existing conclusions for deep quenches.
To this end we introduce an auxiliary Hamiltonian $\mathcal{H}'_{\rm post}(\bf k)$, which is obtained by applying a local rotation $U(\mathbf{k})$ to
the post-quench Hamiltonian $\mathcal{H}_{\rm post}(\bf k)$, i.e., $\mathcal{H}'_{\rm post}(\textbf{k})=U(\textbf{k})\mathcal{H}_{\rm post}(\textbf{k})U^{\dagger}(\textbf{k})$.
Here the local rotation is chosen such that the rotated initial state $\rho'_{0}(\mathbf{k})=U(\mathbf{k})\rho_{0}(\mathbf{k})U^{\dagger}(\mathbf{k})$
becomes fully polarized in this transformed frame: $\gamma_{0}\rho'_{0}(\mathbf{k})=-\rho'_{0}(\mathbf{k})$. Note that the
two states $\rho_{0}(\mathbf{k})$ and $\rho'_{0}(\mathbf{k})$ are in the same topologically trivial regime. Hence the local rotation
$U(\mathbf{k})$ can take a trivial form, which makes the auxiliary Hamiltonian $\mathcal{H}'_{\mathrm{post}}(\mathbf{k})$ be topologically equivalent to the original
post-quench Hamiltonian $\mathcal{H}_{\rm post}(\mathbf{k})$~\cite{Poon2018} (see also Appendix~\ref{App1}). We emphasize that the monopoles are not affected
by this local rotation, and the rotated Hamiltonian thus has the same monopole charges as the original Hamiltonian.
As a result, given the same dBISs, the topological pattens on them should be the same for $\mathcal{H}'_{\rm post}$ and $\mathcal{H}_{\rm post}$.
This argument will be carefully checked in the next subsection.


Before that, we write the local rotation as $U(\mathbf{k})=e^{-\ui\mathbf{u}(\mathbf{k})\cdot\boldsymbol{\gamma}}$,
which rotates the (pseudo)spin around the axis $\hat{\mathbf{u}}\equiv\mathbf{u}/|\mathbf{u}|$ by an angle $2|\mathbf{u}|$. Without loss of generality,
the rotation axis is chosen to be the normal vector of the plane spanned by the $\gamma_0$ axis and the vector field
$\mathbf{h}(\mathbf{k})$. We then have $0\leq 2|\mathbf{u}|<\pi/2$ and $u_0=0$. Under this transformation, the pre-quench Hamiltonian given by
${\cal H}_{\rm pre}(\mathbf{k})=E_0(\mathbf{k})U^\dagger(\mathbf{k})\gamma_0U(\mathbf{k})$ should be
\begin{align}\label{Hpre_u0}
{\cal H}_{\rm pre}(\mathbf{k})&=E_0(\mathbf{k})\cos2|{\bf u}(\mathbf{k})|\gamma_0 \nonumber \\
& \qquad -\ui E_0(\mathbf{k})\frac{\sin2|{\bf u}(\mathbf{k})|}{|{\bf u}(\mathbf{k})|}\sum_{i\neq0}u_i(\mathbf{k})\gamma_0\gamma_i,
\end{align}
where $E_0(\mathbf{k})$ is the band energy of the pre-quench Hamiltonian.
Since the quench is along the $\gamma_0$ axis, the post-quench Hamiltonian takes the form
${\cal H}_{\rm post}(\mathbf{k})={\cal H}_{\rm pre}(\mathbf{k})-\delta m_0\gamma_0$, where $\delta m_0$ denotes the tuned magnetization.
Finally, we obtain the auxiliary  Hamiltonian ${\cal H}'_{\rm post}(\mathbf{k})=E_0(\mathbf{k})\gamma_0-\delta m_0
U(\mathbf{k})\gamma_0U^\dagger(\mathbf{k})$, which gives
\begin{align}\label{Hpost_r0}
{\cal H}'_{\rm post}(\mathbf{k}) & = \left[E_0(\mathbf{k})-\delta m_0\cos2|{\bf u}(\mathbf{k})|\right]\gamma_0 \nonumber \\
& \qquad -\ui\delta m_0\frac{\sin2|{\bf u}(\mathbf{k})|}{|{\bf u}(\mathbf{k})|}\sum_{i\neq0}u_i(\mathbf{k})\gamma_0\gamma_i.
\end{align}
Eqns.~(\ref{Hpre_u0}) and (\ref{Hpost_r0}) are our main results for the proof.


\subsection{Proof and example}~\label{Sec3C}

In this subsection, we utilize the auxiliary Hamiltonian ${\cal H}'_{\rm post}$ to demonstrate the projection approach.
First, it is obvious that the dynamical bulk-surface correspondence holds in the rotated frame.
The dynamical topological pattern on the dBIS of the transformed system, which reflects the winding of the rotated SO field,
characterizes the topology of ${\cal H}'_{\rm post}$.
Second, the time-averaged spin textures in the original or rotated frames vanish
on the same $(d-1)$D hypersurface, resulted from
\begin{align}
{\rm Tr}[\rho_0(\mathbf{k}){\cal H}_{\rm post}(\mathbf{k})]={\rm Tr}[\rho'_0(\mathbf{k}){\cal H}'_{\rm post}(\mathbf{k})]=0
\end{align}
for $\mathbf{k}\in\mathrm{dBISs}$.
Hence the dBISs remain the same under the rotation.
Third, we have shown that the projected directional derivative field ${\bf g}({\bf k})$ defined on dBISs characterizes the SO field [Eq.~(\ref{dgamma})].
Knowing these three facts,
the only thing left to prove is that the winding of the SO field in ${\cal H}_{\rm post}$ is topologically equivalent to
the winding of the rotated SO field in ${\cal H}'_{\rm post}$.
From Eqs.~(\ref{Hpre_u0}) and (\ref{Hpost_r0}), we find that the unit SO field
\begin{align}
\hat{\bf h}_{\rm so}=-\ui(\sin2|{\bf u}|/|{\bf u}|)\sum_{i\neq0}u_i\gamma_0\gamma_i
\end{align}
remains the same under the rotation.
Hence we conclude that
by any shallow quench across the phase transition,
the topological pattern of the dynamical field ${\bf g}({\bf k})$
emerging on the dBIS characterizes the topology of post-quench Hamiltonian ${\cal H}_{\rm post}$,
which manifests itself as a generic dynamical bulk-surface correspondence.

\begin{figure}
\includegraphics[width=0.5\textwidth]{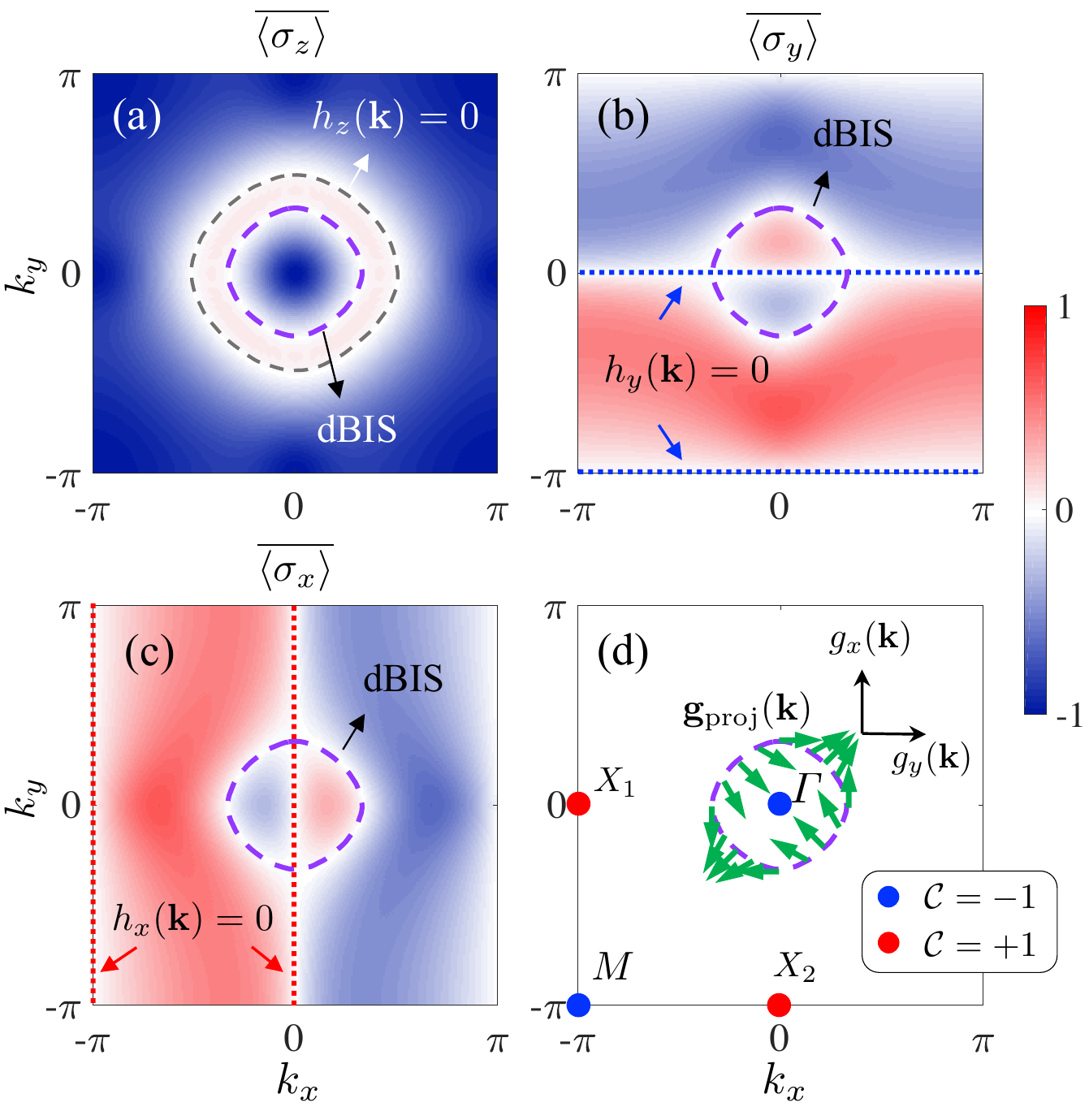}
\caption{Dynamical characterization of 2D QAH model via a shallow quench. (a-c) Time-averaged spin textures $\overline{\langle\sigma_{\alpha}({\bf k})\rangle}$ ($\alpha=x,y,z$).
Here we take $t_{\rm so}=t_0$ and tune $m_z$ from $3t_0$ to $t_0$.
A ring-shape structure (dashed purple) with vanishing polarization emerges in all the spin textures, which characterizes the dBIS.
Besides, another ring (dashed gray) in $\overline{\langle\sigma_z({\bf k})\rangle}$ is the surface with $h_z({\bf k})=0$,
and two dotted lines in $\overline{\langle\sigma_{y/x}({\bf k})\rangle}$ represent $h_{y/x}({\bf k})=0$.
(d) The projected dynamical directional derivative field ${\bf g}_{\rm proj}({\bf k})$ (green arrows), constructed from the spin textures $\overline{\langle\sigma_{x,y}\rangle}$,
reflects the SO field on the dBIS, whose winding classifies the post-quench phase with the Chern number ${\rm Ch}=-1$.
The intersections of the surfaces $h_{x,y}({\bf k})=0$ locate four
monopole charges, with ${\cal C}=-1$ (blue) at $\it\Gamma$ $(0,0)$ and $M$ $(-\pi,-\pi)$, and ${\cal C}=+1$ (red) at $X_{1}$ $(0,-\pi)$ and $X_{2}$ $(-\pi,0)$.
Only the ${\cal C}=-1$ charge at $\it\Gamma$ is enclosed by the dBIS, which also characterizes the topology.
}\label{fig2}
\end{figure}

We further examine if the dynamical spin-texture field
${\bm \Theta}({\bf k})$ given by Eq.~(\ref{thetai}) still works for a shallow quench.
From Eq.~(\ref{gammai}), one can see that the dynamical characterization of
$\overline{\left\langle \gamma_{i}({\bf k})\right\rangle }=0$ for $i\neq0$ exactly determines the location of
monopole charges defined by ${\bf h}_{\rm so}({\bf k})=0$.
It indicates that compared to the dBIS, the monopole charges are immune to the quench depth.
At a monopole ${\bf k}={\bm \varrho}_n$, we have
\begin{align}
\overline{\left\langle \gamma_{i}({\bf k})\right\rangle}\big\vert_{{\bf k}\to{\bm \varrho}_n}\simeq -h_{i}(\bold k)h_0'({\bm \varrho}_n)/h^2_0({\bm \varrho}_n),
\end{align}
where $h_0'({\bf k})\equiv E_0(\mathbf{k})-\delta m_0\cos2|{\bf u}(\mathbf{k})|$.
Since $h'_0({\bm \varrho}_n)=E_0-\delta m_0=h_0({\bm \varrho}_n)$
($|{\bf u}({\bm \varrho}_n)|=0$), we still have $\overline{\left\langle \gamma_{i}({\bf k})\right\rangle}\big\vert_{{\bf k}
\to{\bm \varrho}_n}\simeq-h_{i}(\bold k)/h_0({\bm \varrho}_n)$,
which is same as the result for the deep quench.

Therefore, the dynamical classification theory in Sec.~\ref{Sec2} can be straightforwardly generalized to the shallow quench cases, except that
dBISs take the place of BISs. Here we take the 2D QAH model~\cite{Baozong2018,XJLiu2014,Zhang_book}
as an example, whose Hamiltonian reads ${\cal H}_{\rm QAH}({\bf k})={\bf h}({\bf k})\cdot{\bm \sigma}$,
where ${\bf h}({\bf k})=(t_{\rm so}\sin k_x, t_{\rm so}\sin k_y, m_z-t_0\cos k_x-t_0\cos k_y)$. This model has been
realized in cold atoms~\cite{Wu2016,Sun2017,Sun2018}, and the dynamical characterization
via deep quenches has been analyzed in Ref.~\cite{Zhanglin2018}. The bulk topology can be determined by $m_z$,
with the Chern number ${\rm Ch}=-{\rm sgn}(m_z)$ for $0<|m_z|<2t_0$ and ${\rm Ch}=0$ for $|m_{z}|\geq2t_{0}$.
We take $h_0\equiv h_z$ and the SO field $\bold h_{\rm so}\equiv(h_y,h_x)$. Here the quench process
corresponds to suddenly tuning $m_z$ from $m_{\rm i}$ to $m_{\rm f}$. We numerically calculate the post-quench spin dynamics induced by a
finite magnetization $m_{\rm i}=3t_0$. The time-averaged spin textures $\overline{\langle\sigma_i({\bf k})\rangle}$ ($i=x,y,z$)
are shown in Fig.~\ref{fig2}(a-c), which all show a ring-shape structure, identified as the dBIS.
The dBIS can be moved by the initial magnetization $m_{\rm i}$: When $m_{\rm i}\to\infty$, it should coincide with
the surface $h_z({\bf k})=0$; when $m_{\rm i}\to 2t_0$, it will shrink and disappear.
The dynamical directional derivative field ${\bf g}_{\rm proj}({\bf k})$ is constructed from the spin textures $\overline{\langle\sigma_{x,y}({\bf k})\rangle}$.
For $m_{\rm f}=t_0$, ${\bf g}_{\rm proj}({\bf k})$ winds once along the ring [see (d)], which characterizes the Chern number ${\rm Ch}=-1$.
Moreover, spin textures in (b,c) exhibits two lines with vanishing polarization,
which are the surfaces with $h_y({\bf k})=0$ [for (b)] and $h_x({\bf k})=0$ [for (c)]. These lines have four intersection points
marking the monopoles at ${\it\Gamma}, M$ and $X_{1,2}$ points [(d)], and the charge $\pm1$ at each point can be determined by
the field ${\bm \Theta}({\bf k})$ obtained through Eq.~(\ref{thetai}).
The ring encloses the monopole charge with ${\cal C}=-1$ at ${\it\Gamma}$ point, also giving ${\rm Ch}=-1$.

\section{Generic scheme for a sequence of quenches}~\label{Sec4}
In this section, we consider another dynamical classification scheme, which takes use of a sequence of quenches
along all spin quantization axes.
Each quench is realized by
tuning the constant magnetization $m_i$ in $h_i({\bf k})=m_i+\nu_i({\bf k})$ ($i=0,1,\cdots,d$), with
$\nu_i({\bf k})$ denoting the momentum-dependent term. After quenching $h_i$, we investigate the time-evolved spin texture
of a single spin component $\gamma_0$.
This scheme for deep quenches ($m_i\vert_{t<0}\to\infty$) has been studies in Ref.~\cite{Zhanglong2018}.
The time-averaged $\gamma_0$-polarization reads
\begin{align}
\overline{\langle\gamma_0\rangle}_i=\lim_{T\to\infty}\frac{1}{T}\int_{0}^{T}\ud t\,\mathrm{Tr}\left[\rho_{i}
e^{\ui\mathcal{H}_{\rm post}t}\gamma_{0}e^{-\ui\mathcal{H}_{\rm post}t}\right],
\end{align}
where $\rho_{i}$ denotes the initial state for quenching $h_i$.
In a deep quench case, we have
\begin{align}\label{gamma0i_gammai0}
\overline{\langle\gamma_{0}({\bf k})\rangle}_i=-h_i({\bf k})h_0({\bf k})/E^2({\bf k})=\overline{\langle\gamma_{i}({\bf k})\rangle}_0.
\end{align}
Thus, the dynamical classification can be accomplished in the same procedure as described in Sec.~\ref{Sec2},
but with $\overline{\langle\gamma_{i}\rangle}$ being replaced by $\overline{\langle\gamma_{0}\rangle}_i$~\cite{Zhanglong2018}.
Therefore, the dynamical characterization of BISs is
\begin{align}\label{BIS02}
\overline{\left\langle \gamma_{0}({\bf k})\right\rangle}_i=0,\quad{\rm for}\,\,i=0,1,\cdots,d.
\end{align}
The topological pattern emerging on the BIS is characterized by the dynamical directional derivative field ${\bf f}({\bf k})\equiv(f_1,f_2,\cdots,f_d)$, with
\begin{align}\label{fi}
f_i({\bf k})=-\frac{1}{{\cal N}_k}\partial_{k_\perp}\overline{\left\langle \gamma_{0}\right\rangle}_i.
\end{align}
The location of monopoles is dynamically determined by $\overline{\left\langle \gamma_{0}({\bf k})\right\rangle}_i=0$ for all $i\neq0$,
and the monopole charge is detected by the dynamical spin-texture field ${\bm \Phi}$ with
\begin{align}\label{phii}
\Phi_i({\bf k})\equiv-\frac{{\rm sgn}[h_0({\bf k})]}{{\cal N}_{\bf k}}\overline{\left\langle \gamma_{0}({\bf k})\right\rangle}_i.
\end{align}
In the following, we shall generalize this dynamical classification scheme to the situation with a sequence of shallow quenches.

\subsection{Local rotations}
As in Sec.~\ref{Sec3B},
we also introduce a local transformation $U_i({\bf k})=e^{-\ui {\bf u}_i({\bf k})\cdot{\bm \gamma}}$, with ${\bf u}_i({\bf k})\cdot{\bm \gamma}=\sum_j u_j^{(i)}\gamma_j$, for each quench,
which rotates the initial state $\rho_{i}$ around the axis $\hat{\bf u}_i\equiv{\bf u}_i/|{\bf u}_i|$ by an angle $2|{\bf u}_i|$
to the fully polarized one $\rho'_{i}=U_i({\bf k})\rho_{i}({\bf k})U^\dagger_i({\bf k})$, with $\gamma_{i}\rho'_{i}=-\rho'_{i}$ and
$\gamma_{j\neq i}\rho'_{i}=0$.
We further set $\hat{\bf u}_i$ as the normal vector of the plane spanned by
the vector field ${\bf h}$ and the $\gamma_i$ axis, so that $0\leq2|{\bf u}_i|<\pi/2$ and $u^{(i)}_i=0$.
It can be checked that this transformation does not change the topology.

Note that after each quench, we return to the same post-quench Hamiltonian ${\cal H}_{\rm post}=\sum_jh_j\gamma_j$.
When quenching along $\gamma_i$ axis, we have the pre-quench Hamiltonian ${\cal H}_{\rm pre}^{(i)}={\cal H}_{\rm post}+\delta m_i\gamma_i$,
where $\delta m_i>0$ denotes the tuned magnetization.
The pre-quench vector field should be in the $\gamma_i$ axis after rotation, which gives
\begin{align}\label{Hpre_ri2}
{\cal H}_{\rm pre}^{(i)'}=\left[(h_i+\delta m_i)/\cos2|{\bf u}_i|\right]\gamma_i.
\end{align}
The fact that the band energy is unchanged under rotation leads to the relation $(h_i+\delta m_i)^2/\cos^22|{\bf u}_i|=(h_i+\delta m_i)^2+\sum_{j\neq i}h_j^2$, which yields
\begin{align}
\delta m_i=\cot2|{\bf u}_i|\sqrt{E^2-h_i^2}-h_i,
\end{align}
with $E=(\sum_{j}h_j^2)^{1/2}$.
We use the equality
\begin{align}
U_i\gamma_iU_i^\dagger=&\cos2|{\bf u}_i|\gamma_i+\ui\frac{\sin2|{\bf u}_i|}{|{\bf u}_i|}\sum_{j\neq i}u^{(i)}_j\gamma_i\gamma_j,
\end{align}
and obtain the rotated post-quench Hamiltonian 
\begin{align}
{\cal H}_{\rm post}^{(i)'}=&{\cal H}_{\rm pre}^{(i)'}-\delta m_iU_i\gamma_iU^\dagger_i\nonumber\\
=&\frac{h_i+\delta m_i\sin^22|{\bf u}_i|}{\cos2|{\bf u}_i|}\gamma_i-\ui\delta m_i\frac{\sin2|{\bf u}_i|}{|{\bf u}_i|}\sum_{j\neq i}u^{(i)}_j\gamma_i\gamma_j.
\end{align}
Thus, we have ${\rm Tr}[\rho_i {\cal H}_{\rm post}]={\rm Tr}[\rho'_i {\cal H}^{(i)'}_{\rm post}]=-\widetilde{h}_i$, where we have denoted
\begin{align}\label{hitilde}
\widetilde{h}_i&\equiv(h_i+\delta m_i\sin^22|{\bf u}_i|)/\cos2|{\bf u}_i|\nonumber\\
&=h_i\cos2|{\bf u}_i|+\sin2|{\bf u}_i|\sqrt{E^2-h_i^2}.
\end{align}

\begin{figure}
\includegraphics[width=0.5\textwidth]{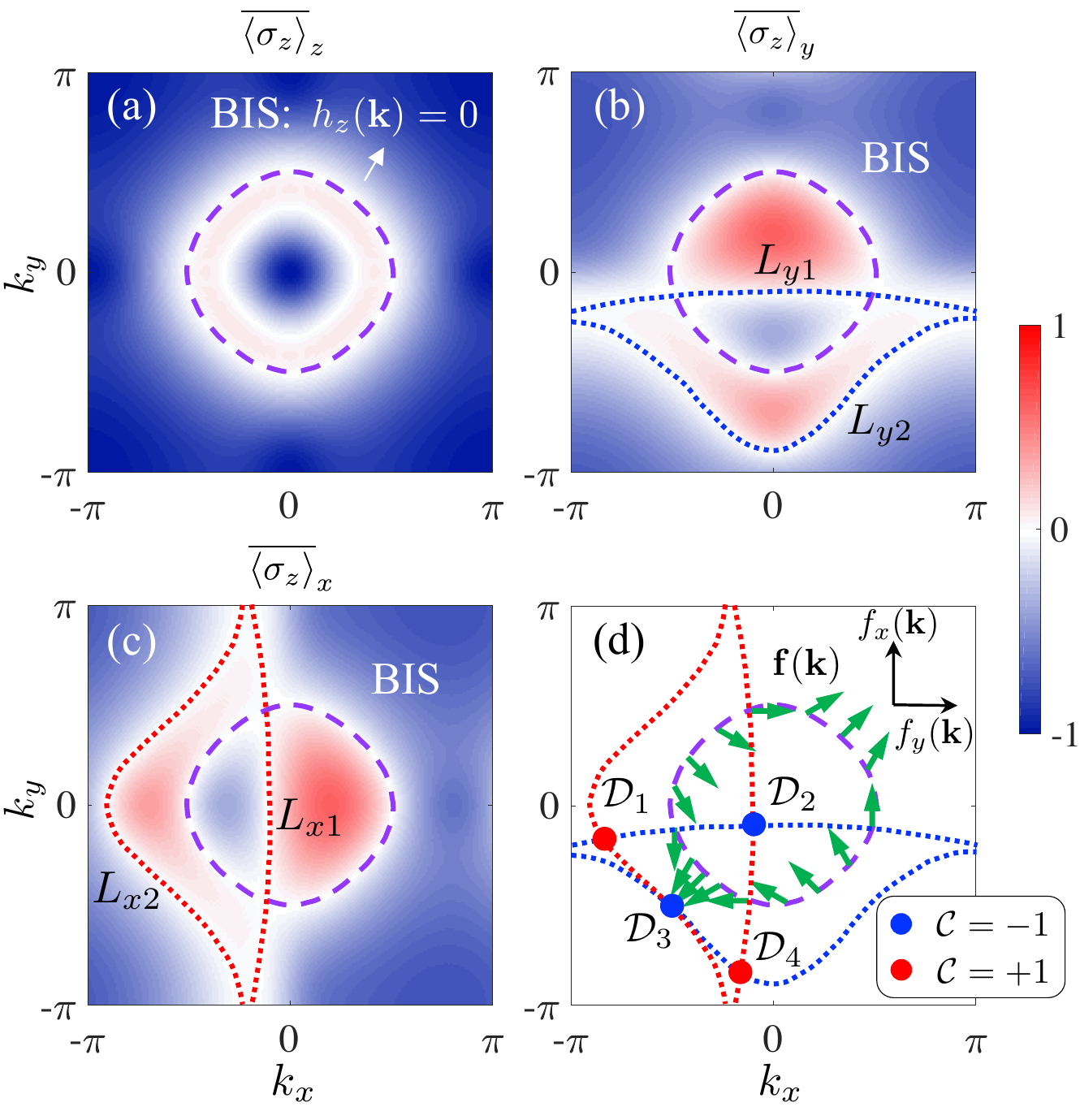}
\caption{Dynamical characterization of 2D QAH model following a series of shallow quenches.
(a-c) Time-averaged spin textures $\overline{\langle\sigma_{z}({\bf k})\rangle}_\alpha$ after quenching $h_\alpha$ was ($\alpha=x,y,z$).
Each quench corresponds to tuning $m_\alpha$ from $m_\alpha^{\rm i}$ to $m_\alpha^{\rm f}$.
Here we take $t_{\rm so}=t_0$ and set $m^{\rm i}_{x,y,z}=3t_0$, $m^{\rm f}_{x,y}=0$ and $m^{\rm f}_{z}=t_0$.
A ring-shape structure emerging in all the spin textures characterizes the BIS with $h_z({\bf k})=0$.
Besides, there are also two curves $L_{y1,y2}$ (or $L_{x1,x2}$)
with vanishing polarization in $\overline{\langle\sigma_z({\bf k})\rangle}_y$ (or $\overline{\langle\sigma_z({\bf k})\rangle}_x$).
(d) The dynamical directional derivative field ${\bm f}({\bf k})$ (green arrows), constructed from the spin textures $\overline{\langle\sigma_z\rangle}_{x,y}$,
is plotted on the BIS, whose winding classifies the post-quench phase with the Chern number ${\rm Ch}=-1$.
The intersections of the surfaces $L_{x1,x2}$ and $L_{y1,y2}$ locate four dynamical charges, with
${\cal C}=+1$ (red) at ${\cal D}_{1,4}$ and ${\cal C}=-1$ (blue) at ${\cal D}_{2,3}$.
Only the ${\cal C}=-1$ charge is enclosed by the BIS, which also characterizes the topology.
}\label{fig3}
\end{figure}

\begin{figure*}
\includegraphics[width=\textwidth]{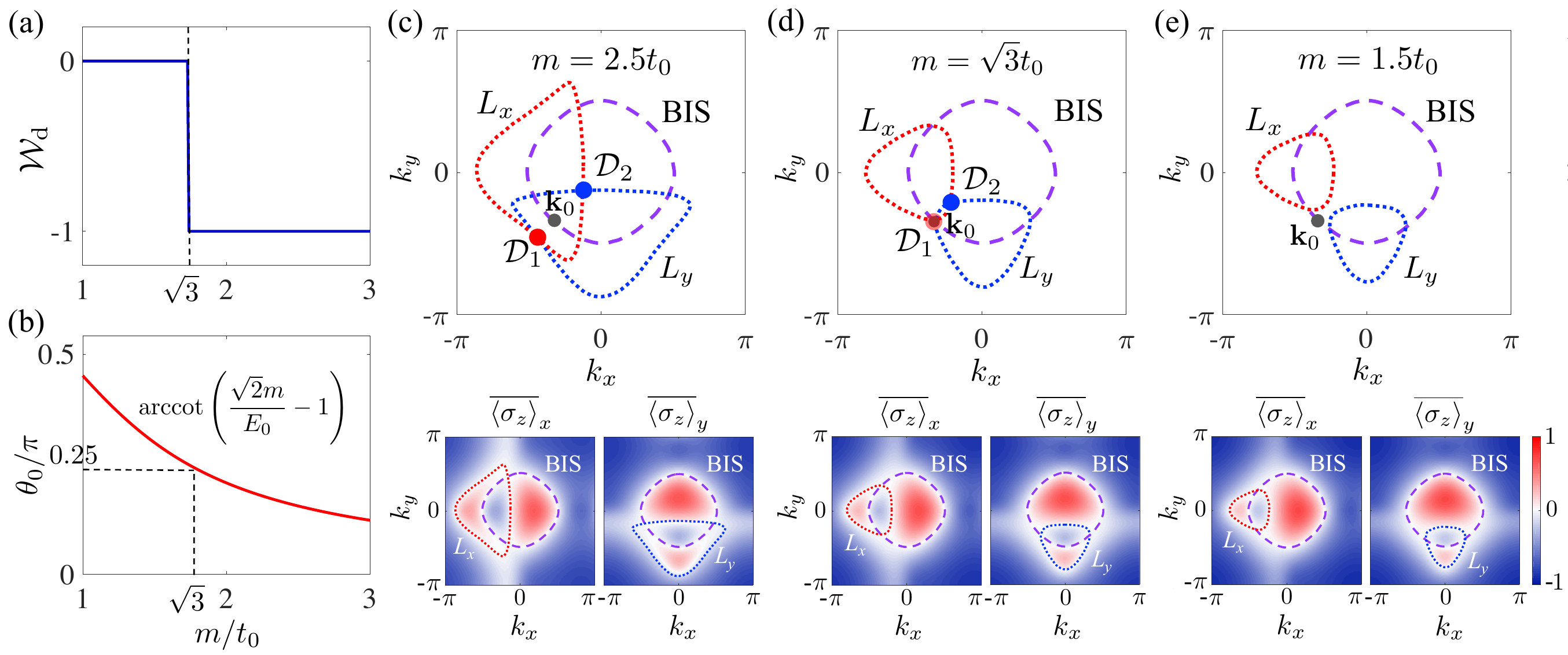}
\caption{Emergent dynamical topological transition in 2D QAH model.
(a) The dynamical topological invariant ${\cal W}_{\rm d}$ measured as the total charges enclosed by the BIS
for various magnetization $m$ [see (c-e) for example]. When $m>\sqrt{3}t_0$, the topological
number ${\cal W}_{\rm d}$ correctly characterizes the post-quench phase.
(b) The deviation angle $\theta_0$  from the polarization axis at the momentum ${\bf k}_0$ [see (c-e)]
as a function of the magnetization $m$, with $E_0\equiv E({\bf k}_0)=t_{\rm so}\sqrt{6}/2$.
At the transition point $m=\sqrt{3}t_0$, we have $\theta_0=45^\circ$.
(c-e) Emergent dynamical topology exhibited by time-averaged spin textures
$\overline{\langle\sigma_z\rangle}_{x,y}$ at $m=2.5t_0$ (c), $\sqrt{3}t_0$ (d), and $1.5t_0$ (e).
Two closed curves $L_{x}$ and $L_{y}$ are the surfaces with vanishing polarization in $\overline{\langle\sigma_z\rangle}_{x}$
and $\overline{\langle\sigma_z\rangle}_{y}$, respectively (lower panel), and their intersections denote
two dynamical topological charges ${\cal D}_{1,2}$ (upper panel).
When $m>\sqrt{3}t_0$,  only one charge ${\cal D}_2$ is enclosed by the BIS (c); when $m\to\sqrt{3}t_0$, the charge ${\cal D}_{1}$ moves
to the momentum ${\bf k}_0$ on the BIS (d); when $m<\sqrt{3}t_0$, the enclosed total charge is zero (e).
In all our calculations, we set $t_{\rm so}=t_0$, $m^{\rm f}_{x,y}=0$ and $m^{\rm f}_{z}=t_0$.
}\label{fig4}
\end{figure*}

\subsection{Validity condition}

Generally, we have
\begin{align}
\overline{\langle\gamma_{0}\rangle}_i=h_0{\rm Tr}[\rho_i{\cal H}_{\rm post}]/E^2=-h_0\widetilde{h}_i/E^2.
\end{align}
It is easily seen that the dynamical characterization of $\overline{\langle\gamma_{0}({\bf k})\rangle}_i=0$ for all $i=0,1,\cdots,d$
exactly characterizes the BIS where $h_0({\bf k})=0$ even when the quenches are all shallow.
Further calculations yield
\begin{align}
\partial_{k_\perp}\overline{\left\langle \gamma_{0}\right\rangle}_i=-\widetilde{h}_i({\bf k})/E^2({\bf k})
\end{align}
on the BIS $h_0({\bf k})=0$.
Hence we have $f_i({\bf k})\vert_{{\bf k}\in {\rm BIS}}\simeq \widetilde{h}_i({\bf k})$.
We then check the topological pattern of $(\widetilde{h}_1, \widetilde{h}_2,\dots,\widetilde{h}_d)$ on BISs to examine if the directional derivative field ${\bf f}({\bf k})$
can characterize the post-quench topology.

We find that when only considering the first term in Eq.~(\ref{hitilde}), the transformed field, with $\widetilde{h}_i=h_i\cos2|{\bf u}_i|$, preserves the topology of the SO field.
It can be checked by examining the lower-dimensional Hamiltonian
${\cal H}_g=g{\cal H}_{\rm so}+(1-g){\cal H}_{\rm ft}$,
with ${\cal H}_{\rm so}\equiv\sum_{i>0}h_i\gamma_i$ and ${\cal H}_{\rm ft}\equiv\sum_{i>0}h_i\cos2|{\bf u}_i|\gamma_i$,
which is gapped for $0\leq g\leq1$. It means that
there is a smooth path connecting ${\cal H}_{\rm so}$ ($g=1$) to  ${\cal H}_{\rm ft}$ ($g=0$);
thus ${\cal H}_{\rm ft}$ is topologically equivalent to ${\cal H}_{\rm so}$.
In contrast, the second term in Eq.~(\ref{hitilde}) can change the topology
if there is no momentum on BISs making
\begin{align}\label{hi_condition}
\widetilde{h}_i({\bf k})<0,\quad{\rm for\,\,all}\,\,i>0.
\end{align}
Note that Eq.~(\ref{hi_condition}) is a necessary but not a sufficient condition for a valid dynamical characterization.
This condition holds only at ${\bf k}$ where $h_{i>0}({\bf k})<0$, and requires
\begin{align}\label{tan_condition}
\tan2|{\bf u}_i({\bf k})|<-\frac{h_i({\bf k})}{\sqrt{E^2({\bf k})-h_i^2({\bf k})}}.
\end{align}
Since $\tan2|{\bf u}_i|=\sqrt{E^2-h_i^2}/(h_i+\delta m_i)$
and $h_i+\delta m_i>0$, we have
\begin{align}\label{hi_less}
h_i({\bf k})<-E({\bf k})^2/\delta m_i.
\end{align}
Therefore, in order to correctly characterize the post-quench topology,
there must exist momenta on BISs satisfying 
Eq.~(\ref{hi_less}) for all $i>0$.

We then examine if the dynamical characterization of topological charges still works.
The locations of charges are dynamically determined by
\begin{align}
\widetilde{h}_i({\bf k})=0,\quad{\rm for\,\,all}\,\,i>0.
\end{align}
These charges are obviously different from those monopole charges defined by ${\bf h}_{\rm so}({\bf k})=0$, and hence dubbed as {\it dynamical} charges.
Moreover, the dynamical spin-texture field ${\bm \Phi}$ defined in Eq.~(\ref{phii}) characterizes these movable dynamical charges.
We have proved that the winding of ${\bf f}({\bf k})$ on BISs can classify the bulk topology only when the condition (\ref{hi_less}) 
is satisfied.
It is then expected that under the same condition, the total dynamical charges enclosed by BISs can be used for the dynamical characterization.

\subsection{A typical case and the dynamical topological transition}

We consider a simple but typical case in experimental realization~\cite{Zhanglong2018,TopoChargeExp} that
$\delta m_i=m$ for all $i>0$.
Generally, there is at least one momentum ${\bf k}_0$ on BISs where $h_1({\bf k}_0)=h_2({\bf k}_0)=\cdots=h_d({\bf k}_0)=-E_0/\sqrt{d}$, with $E_0\equiv E({\bf k}_0)$.
We find that in this situation, Eq.~(\ref{hi_less}) with ${\bf k}={\bf k}_0$ reduces to
\begin{align}\label{m0_condition}
m>E_0\sqrt{d},
\end{align}
which is the necessary and sufficient condition for a valid dynamical characterization (see Appendix~\ref{App2}).
Alternatively, we define $\theta_0\equiv2|{\bf u}_i({\bf k}_0)|$ and have
\begin{align}\label{theta0_condition}
\theta_0<{\rm arccot}\sqrt{d-1},
\end{align}
which is equivalent to Eq.~(\ref{m0_condition}).
When $d=1$, we have $\theta_0<90^\circ$, which means that in one dimensions, the dynamical characterization is valid for any shallow quenches.
When $d=2$, the criterion $\theta_0<45^\circ$ delimits the range of $m$ (see the following example).
When $d\to\infty$, Eq.~(\ref{theta0_condition}) reduces to $\theta=0^\circ$;
the dynamical topological number correctly characterizes the topology when the quenches are extremely deep.

For illustration, we still consider the 2D QAH model as in Sec.~\ref{Sec3C}, but with
${\bf h}({\bf k})=(m_x+t_{\rm so}\sin k_x, m_y+t_{\rm so}\sin k_y, m_z-t_0\cos k_x-t_0\cos k_y)$.
The dynamical characterization via a sequence of deep quenches has been analyzed in Ref.~\cite{Zhanglong2018}.
We take $h_0\equiv h_z$, and the quench is performed by suddenly varying $m_\alpha$ ($\alpha=x,y,z$) from $m^{\rm i}_\alpha$ to $m^{\rm f}_\alpha$
($m=m^{\rm i}_{x/y}-m^{\rm f}_{x/y}$).
Note that only $\sigma_z$-component of the spin polarization needs to be measured, which is well achievable in cold atom experiments.
Here we set $m^{\rm f}_{x,y}=0$ and $m^{\rm f}_{z}=t_0$.
Hence, the momentum ${\bf k}_0$ on the BIS which makes $h^{\rm f}_x({\bf k}_0)=h^{\rm f}_y({\bf k}_0)<0$ lies at
$k_x=k_y=-\pi/3$ with $E_0=t_{\rm so}\sqrt{6}/2$.

In Fig.~\ref{fig3}, we show the dynamical characterization of the post-quench phase with $m=3t_0$.
We numerically calculate the spin dynamics after each quench.
The time-averaged spin textures $\overline{\langle\sigma_z({\bf k})\rangle}_\alpha$ ($\alpha=x,y,z$)
are shown in Fig.~\ref{fig3}(a-c), where a ring-shape structure emerges and identified as the BIS where $h_z({\bf k})=0$.
The dynamical directional derivative field ${\bf f}({\bf k})$ is constructed from the spin textures $\overline{\langle\sigma_z({\bf k})\rangle}_{x,y}$.
The winding of ${\bf f}({\bf k})$ along the BIS characterizes the Chern number of the post quench phase [Fig.~\ref{fig3}(d)].
Moreover, spin textures in (b,c) also exhibits two curves with vanishing polarization. These curves have four intersection points,
which mark the locations of dynamical topological charges.
These charges can be characterized by the dynamical field ${\bm \Phi}({\bf k})$ through Eq.~(\ref{phii}).
It shows that only one ${\cal C}=-1$ charge is enclosed by the ring, giving the Chern number ${\rm Ch}=-1$.

The criterion (\ref{m0_condition}) indicates that there exists a transition of the emergent topology exhibited by $d$-axial quench dynamics, associated with dynamical charges moving across BISs.
In the parameter regime satisfying Eq.~(\ref{m0_condition}), the dynamical topological invariant measured from time-averaged spin textures
directly characterizes the post-quench system.
In other regimes, the quench dynamics may exhibit a different emergent topology. 
In this case, the topology of the post-quench phase is characterized by the emergent topological invariant plus (minus) the total charges moving outside (inside) the region $V_-$.
Figure \ref{fig4} shows the emergent dynamical topological transition in the QAH model.
In the picture of topological charges [Fig.~\ref{fig4}(c-e)], the measured dynamical topological variant ${\cal W}_d$ has two distinct regimes
with respect to the magnetization $m$, with the transition point being $m_{\rm t}=\sqrt{3}t_0$ [see (a)].
When $m\to m_t$, one dynamical charge moves to the momentum ${\bf k}_0$ and then crosses the BIS [see (d)], making the total charges enclosed by the BIS changed.
The angle $\theta_0$ is also shown in Fig.~\ref{fig4}(b) as a function of $m$, describing the deviation from full polarization.

\section{Discussion and conclusion}\label{Sec5}

We have examined two generic dynamical classification schemes---Scheme I uses a single quench from a trivial regime
and Scheme II employs a sequence of quenches with respect to all spin axes.
The central idea of our non-equilibrium theories
is to characterize topological phases by dynamical topological invariants
emerging from unitary dynamics of a topologically trivial state.
This idea is embodied in the dynamical bulk-surface correspondence,
which is not only of theoretical significance, but also has practical benefits for experimental realization~\cite{Sun2018}.
Former studies~\cite{Zhanglin2018,Zhanglong2018} all focused on a simple case that the initial trivial state is fully polarized.
We broaden the application of the dynamical characterization by considering shallow quenches.

From Eq.~(\ref{gamma0i_gammai0}), one can see that on the theoretic level, Schemes I and II are equivalent in the deep quench limit.
In this work, we have shown that Schemes I and II are generally different when performing shallow quenches.
Compared to Schemes I, which always works for an incompletely polarized state, Scheme II
requires a constraint on the quench depth, i.e., on the polarization of the initial trivial phase, since various degrees of freedom
are involved in the characterization. The formulas (\ref{m0_condition}) and (\ref{theta0_condition}) imply that
the criterion also depends on the system dimension.

In summary, we have demonstrated the validity of two non-equilibrium classification schemes through local transformations.
Our work shows that the dynamical bulk-surface correspondence holds for a wide range of quench ways.
Being of high feasibility, the present study can be immediately applied in current experiments~\cite{TopoChargeExp}.

\begin{acknowledgments}
This work was supported by the National Key R\&D Program of China (Project No. 2016YFA0301604),
National Natural Science Foundation of China (Grants No. 11574008, No. 11761161003, and No. 11825401), and the
Strategic Priority Research Program of Chinese Academy of Science (Grant No. XDB28000000).
\end{acknowledgments}

\begin{appendix}

\section{Proof of the continuous transformation}\label{App1}
In this appendix, we prove that the constructed transformation $U({\bf k})$ can continuously connect the Hamiltonian
${\cal H}$  to ${\cal H}'=U({\bf k}){\cal H}U^\dagger({\bf k})$ without change of topology.
Generically, we need to examine if the Hamiltonian ${\cal H}^G\equiv g{\cal H}+(1-g){\cal H}'$ is always gapped for $0\leq g\leq1$.

Without loss of generality, we consider the rotation in Sec.~\ref{Sec3B}, where $U(\mathbf{k})=e^{-\ui\mathbf{u}(\mathbf{k})\cdot\boldsymbol{\gamma}}$
with $0\leq 2|\mathbf{u}|<\pi/2$ and $u_0=0$. We check the topology change of pre- and post-quench Hamiltonians under the rotation.
Since
\begin{align}
U\gamma_{0}U^{\dagger} =\cos2|\mathbf{u}|\gamma_{0}+\ui\frac{\sin2|\mathbf{u}|}{|\mathbf{u}|}\sum_{j\neq0}u_{j}\gamma_{0}\gamma_{j},
\end{align}
we write ${\cal H}_{\rm pre}'({\bf k})=E_0({\bf k})\gamma_0$ and ${\cal H}_{\rm pre}({\bf k})$ takes the form in Eq.~(\ref{Hpre_u0}). We then obtain
\begin{align}
{\cal H}_{g1}&=g{\cal H}_{\rm pre}+(1-g){\cal H}'_{\rm pre}\nonumber\\
&=E_0\left[\left(1-g+g\cos2|{\bf u}|\right)\gamma_0-\ui g\frac{\sin2|{\bf u}|}{|{\bf u}|}\sum_{i\neq0}u_i\gamma_0\gamma_i\right],
\end{align}
which gives
\begin{align}
{\cal H}^2_{g1}=E_0^2\left[(1-g)^2+2(1-g)g\cos2|{\bf u}|+g^2\right]>0.
\end{align}
To examine the topological equivalence between ${\cal H}_{\rm post}$ and
${\cal H}'_{\rm post}$, we need to check
the Hamiltonian
\begin{align}
{\cal H}_{g2}&=g{\cal H}_{\rm post}+(1-g){\cal H}'_{\rm post}\nonumber\\
&=\{E_0(1-g+g\cos2|{\bf u}|)-\delta m_0[g+(1-g)\cos2|{\bf u}|)]\}\nonumber\\
&\gamma_0-\ui [gE_0+(1-g)\delta m_0]\frac{\sin2|{\bf u}|}{|{\bf u}|}\sum_{i\neq0}u_i\gamma_0\gamma_i.
\end{align}
We then have
\begin{align}
{\cal H}_{g2}^2
=&\left[(1-g)^2+2g(1-g)\cos2|{\bf u}|+g^2\right]\nonumber\\
&\times\left(E_0^2+\delta m_0^2-2\,\delta m_0E_0\cos2|{\bf u}|\right)>0.
\end{align}
Therefore both the pre- and post-quench Hamiltonians are topologically unchanged under the rotation $U({\bf k})$.
This demonstration can be also applied to the local rotations introduced in the dynamical characterization by
a sequence of shallow quenches.

\section{The necessary and sufficient condition}\label{App2}

In this appendix, we derive the necessary and sufficient condition for the dynamical characterization via a sequence of quenches with $\delta m_i=m$ ($i>0$).
We first define three lower-dimensional Hamiltonians ${\cal H}_{\rm so}\equiv\sum_{i>0}h_i\gamma_i$,
${\cal H}_{\rm rot}\equiv\sum_{i>0}\widetilde{h_i}\gamma_i$,
and ${\cal H}_{\rm ft}\equiv\sum_{i>0}h_i\cos2|{\bf u}_i|\gamma_i$.
It is easily seen that the Hamiltonian ${\cal H}_{g3}=g{\cal H}_{\rm so}+(1-g){\cal H}_{\rm ft}$ is always gapped for $0\leq g\leq1$.
To examine the topological equivalence between ${\cal H}_{\rm so}$ and ${\cal H}_{\rm rot}$, we examine
the Hamiltonian ${\cal H}_{g4}=g{\cal H}_{\rm ft}+(1-g){\cal H}_{\rm rot}$,
and have
\begin{align}
{\cal H}_{g4}^2
=&\sum_{i>0}\frac{\left[h_i(h_i+m)+(1-g)(E^2-h_i^2)\right]^2}{E^2+2mh_i+m^2}.
\end{align}
Suppose that there is a momentum ${\bf k}$ that makes the Hamiltonian ${\cal H}_{g4}$ gapless at $g=g_0$ ($0<g_0<1$).
We then have
\begin{align}
E^2({\bf k})=h_i^2({\bf k})-\frac{h_i({\bf k})[h_i({\bf k})+m]}{1-g_0}
\end{align}
for all $i>0$.
Hence the only possible gapless point is ${\bf k}={\bf k}_0$ where $h_1({\bf k}_0)=h_2({\bf k}_0)=\cdots=h_d({\bf k}_0)=-E_0/\sqrt{d}$, with $E_0\equiv E({\bf k}_0)$.
In order to make ${\cal H}_{g4}$ always gapped, we should ensure
$E_0^2<-mh_i({\bf k}_0)$,
which leads to
\begin{align}
m>E_0\sqrt{d}.
\end{align}
This is the validity condition of our dynamical classification theory.
We also have
\begin{align}
\tan\theta_0=\frac{\sqrt{E_0^2-h_i^2({\bf k}_0)}}{h_i({\bf k}_0)+m}<\frac{1}{\sqrt{d-1}},
\end{align}
which gives Eq.~(\ref{theta0_condition}).

\end{appendix}


\noindent

\end{document}